# Valley-coherent quantum anomalous Hall state in AB-stacked MoTe$_2$/WSe$_2$ bilayers

Zui Tao[1*], Bowen Shen[1*], Shengwei Jiang[2*], Tingxin Li[1*], Lizhong Li[1], Liguo Ma[1], Wenjin Zhao[3], Jenny Hu[4], Kateryna Pistunova[4], Kenji Watanabe[5], Takashi Taniguchi[5], Tony F. Heinz[4,6], Kin Fai Mak[1,2,3**], and Jie Shan[1,2,3**]

[1]School of Applied and Engineering Physics, Cornell University, Ithaca, NY, USA
[2]Laboratory of Atomic and Solid State Physics, Cornell University, Ithaca, NY, USA
[3]Kavli Institute at Cornell for Nanoscale Science, Ithaca, NY, USA
[4]Departments of Physics and Applied Physics, Stanford University, Stanford, CA, USA
[5]National Institute for Materials Science, 1-1 Namiki, 305-0044 Tsukuba, Japan
[6]SLAC National Accelerator Laboratory, Menlo Park, CA, USA

[*]These authors contributed equally.
[**]Email: jie.shan@cornell.edu; kinfai.mak@cornell.edu

**Abstract**
Moiré materials provide fertile ground for the correlated and topological quantum phenomena. Among them, the quantum anomalous Hall (QAH) effect, in which the Hall resistance is quantized even under zero magnetic field, is a direct manifestation of the intrinsic topological properties of a material and an appealing attribute for low-power electronics applications. The QAH effect has been observed in both graphene and transition metal dichalcogenide (TMD) moiré materials. It is thought to arise from the interaction-driven valley polarization of the narrow moiré bands. Here, we show surprisingly that the newly discovered QAH state in AB-stacked MoTe$_2$/WSe$_2$ moiré bilayers is not valley-polarized but valley-coherent. The layer- and helicity-resolved optical spectroscopy measurement reveals that the QAH ground state possesses spontaneous spin (valley) polarization aligned (anti-aligned) in two TMD layers. In addition, saturation of the out-of-plane spin polarization in both layers occurs only under high magnetic fields, supporting a canted spin texture. Our results call for a new mechanism for the QAH effect and highlight the potential of TMD moiré materials with strong electronic correlations and spin-orbit interactions for exotic topological states.

**Main text**

Narrow bands with nontrivial topology are believed to be the key to realizing interaction-driven topological states (*1-3*). Moiré superlattices formed in van der Waals bilayers with a small twist angle or lattice mismatch can induce narrow moiré bands (*1, 4-6*). These narrow bands enhance the importance of electronic correlations, as manifested in the emergence of superconductivity (*4, 7*) and a set of correlated insulating states in graphene and TMD moiré materials (*1, 4-6*). Topological states have also been proposed in these materials based on valley contrast physics, that is, moiré bands possess opposite Chern numbers in the K and K' valleys of the Brillouin zone (*1, 2, 8-13*). In particular, the QAH effect has been observed in graphene-based moiré systems (*14-19*), and is consistent with a fully valley-polarized ground state as a result of the interaction-driven valley symmetry breaking (*1, 2, 8, 10-13*). A similar mechanism has been put forward for AA-stacked (near-0°-aligned) TMD homobilayers such as $MoTe_2$ and $WSe_2$ (*9, 20*). Unlike graphene, TMD monolayers have broken inversion symmetry, and possess strong Ising spin-orbit interactions and spin-valley locking (*21, 22*). Because of the large spin splitting at the monolayer valence band maximum, the low-energy valence states of the bilayer at the K (or K') valley can be described by two bands, one from each layer. Valley Chern bands with opposite Chern numbers can form from a topologically nontrivial interlayer tunneling structure that has the moiré period (*9, 20*). A topological Kane-Mele model is realized with the Wannier orbitals of the two layers forming a honeycomb lattice.

Recently, the QAH effect was observed in AB-stacked (near-60°-aligned) $MoTe_2/WSe_2$ moiré bilayers (*23*). In the absence of an out-of-plane electric field, $E$, the bands are expected to be topologically trivial (*24, 25*); the system is a Mott insulator with one hole per moiré unit cell that resides in $MoTe_2$. A large electric field reduces the band offset between the two layers, inverts the bands, and drives a topological phase transition to a QAH insulator (also referred to as Chern insulator). The experimental result is quite surprising: although AB-stacked $MoTe_2/WSe_2$ moiré bilayers can be described by a two-band model on a honeycomb lattice (*24, 25*) as in the case of AA-stacked TMD moiré homobilayers (*9, 20*) (Fig. 1A), the two bands have opposite spins, and interlayer tunneling is spin-forbidden (Fig. 1B). A variety of theories has been proposed to explain the experimental result (*24-32*). It was pointed out that valley Chern bands could still form from a strain pseudo-magnetic field (*26*) or from interlayer tunneling when the out-of-plane spin is no longer conserved (*24*); both can be induced by lattice reconstruction in moiré bilayers. These theories naturally predict a (fully or partially) valley-polarized QAH ground state (*24-29, 32*).

In this study, we directly probe the magnetic properties of the QAH state in AB-stacked $MoTe_2/WSe_2$ moiré bilayers by layer-resolved and helicity-resolved optical spectroscopy. We show that the holes are distributed in both layers, and contrary to the theoretical predictions (*24-29, 32*), the QAH ground state is not valley-polarized but is consistent with a valley-coherent state, i.e. hybridization of states from different layers and different valleys near the Fermi level, thus opening a correlated charge gap. Theoretical studies beyond the current frame of non-interacting valley states may be required to understand the experimental results (*30, 31*).

**Optical selection rules**
The current measurement relies on the unique optical selection rules for intralayer excitations in TMDs (*21, 22*) (Fig. 1C). In the presence of doping, the optical excitations are dominated by polarons (or charged excitons) instead of neutral excitons (*33, 34*); particularly, the attractive polaron is a bound state of a photo-excited exciton in one valley and charge excitations in the other valley due to the Pauli exclusion principle. The attractive polaron with exciton in the K (K') valley is exclusively coupled to the left (right) circularly polarized light, $\sigma^+$ ($\sigma^-$), in the MoTe$_2$ layer. Magnetic circular dichroism (MCD) emerges when the layer is spin-valley polarized. In the extreme case of doped holes occupying only one spin-valley state, the attractive polaron response vanishes for one of the helicities (*22, 33, 35*). Furthermore, because of the AB stacking structure, the optical selection rules in the WSe$_2$ layer are reversed (identical) for valley (spin). Hence, if the attractive polaron response in two layers is dominated by excitation of opposite helicities, the holes occupy opposite spin states, and the bilayer is valley-polarized (Fig. 1D). Conversely, if the response is dominated by identical helicities, the holes occupy same spin states, and the bilayer is in a superposition of two valley states (Fig. 1E). The optical selections have been independently verified in separate MoTe$_2$ and WSe$_2$ monolayers polarized by a large out-of-plane magnetic field (Supplementary Fig. 1).

**QAH effect and topological phase transition**
We investigate dual-gated devices of AB-stacked MoTe$_2$/WSe$_2$ moiré bilayers with doping density of one hole per moiré unit cell at 1.6 K unless otherwise specified. Details on the device fabrication and measurements are provided in Supplementary Materials. Figure 2A (left) shows the transport data at $E \approx 0.693$ V/nm that supports the QAH effect. Particularly, under zero magnetic field the Hall resistance, $R_{xy}$, is nearly quantized at the value of $h/e^2$ ($h$ and $e$ denoting the Planck's constant and elementary charge, respectively), and the longitudinal resistance, $R_{xx}$, is negligible. Both $R_{xx}$ and $R_{xy}$ display a hysteretic magnetic-field dependence with a small coercive field that depends on the field scan rate and sample details.

Figure 2B (middle) is the corresponding helicity-resolved optical reflectance contrast (RC) spectrum under zero magnetic field. The top panel shows the RC for a charge neutral device as a reference. The low- and high-energy spectra are dominated by the response of MoTe$_2$ and WSe$_2$, respectively. The spectral lineshape depends on the multilayer interference effect in the device. In the QAH state, RC is dominated by polarons in both layers, including the attractive and repulsive polarons that are red- and blue-shifted, respectively, from the neutral excitons (the weak repulsive polaron in the heavily hole-doped MoTe$_2$ is outside the detector range). This indicates that both layers are hole-doped. In addition, there is spontaneous spin-valley polarization with nonzero MCD in both layers. The MCD spectrum (bottom panel) is defined as $\frac{I^- - I^+}{I^- + I^+}$, where $I^-$ and $I^+$ denote the reflection intensity of the $\sigma^-$ and $\sigma^+$ light, respectively. We use the maximum MCD of the attractive polaron feature (averaged over the narrow shaded spectral window) to characterize the spin-valley polarization in each layer (WSe$_2$ shown in the right panel of Fig. 2A). Figure 2A shows that the quantized Hall resistance and the spontaneous spin-valley polarization are correlated.

The correlation between the QAH effect and the spontaneous spin-valley polarization is also supported by the electric-field dependence of the transport and optical properties under zero magnetic field (Fig. 2C). When the electric field exceeds a critical value, $E_c \approx 0.687$ V/nm (dashed line), $R_{xx}$ drops by orders of magnitude to about 1 kΩ; and concurrently, $R_{xy}$ increases from zero to near $h/e^2$ (Fig. 2C top panel). This is fully consistent with a recent study that reports an electric-field-tuned topological quantum phase transition from a Mott insulator to a QAH insulator (*23*). Correlated, the spontaneous spin-valley polarization of each layer increases from zero to a finite value at $E_c$; but unlike $R_{xy}$, it does not exhibit a plateau (Fig. 2C middle panel). With further increase of $E$, $R_{xy}$ and the spontaneous polarization in both layers decrease, signaling a departure from the QAH state.

The layer-resolved RC can also probe the charge distribution in the bilayer (Fig. 2C bottom panel). At small $E$'s, the helicity-unresolved RC spectrum of the WSe$_2$ layer is dominated by a single neutral exciton resonance, indicating a charge neutral layer; at large $E$'s, the spectrum consists of two polaron resonances, indicating a doped layer. The transition, determined from the emergence of the polarons, also occurs at $E_c$. In contrast, the MoTe$_2$ layer is doped for the entire electric-field range (Supplementary Fig. 2). The correlation between the emergence of the QAH state and onset of charge transfer from the MoTe$_2$ layer to the WSe$_2$ layer further supports the picture that for $E < E_c$ holes reside in MoTe$_2$ and the system is a Mott insulator, and for $E > E_c$ the QAH state arises from electric-field-tuned band inversion and mixing (*24, 25*).

**Spontaneous polarization alignment**
Next we examine the relative alignment of the spontaneous spin (valley) polarization in the bilayer in the QAH state. Figure 3A exhibits the magnetic-field dependence of the helicity-resolved reflection contrast $RC^-$ (top) and $RC^+$ (bottom). Figure 3B is the MCD spectrum extracted from Fig. 3A. It shows an abrupt sign change near zero field. Magnetic hysteresis can be observed in MCD for a small field range (Supplementary Fig. 3). In Fig. 3A, the electric field (≈ 0.693 V/nm) was chosen from the middle of the QAH phase; the magnetic field, $B$, was swept from – 8 T to 8 T; and the two helicity-resolved spectra were measured together at each $B$. In addition, since the QAH state disperses in density and magnetic field following the Streda relation (*23*), we tune the doping density for each field accordingly in order to remain in the QAH state.

A linecut of Fig. 3A,B at $B = 0$ (in backward scan) is displayed in Fig. 2B. In both layers, the attractive polaron responds mostly to the $\sigma^-$ excitation. (The repulsive polaron in WSe$_2$ responds mostly to the $\sigma^+$ excitation; this is consistent with the reported opposite optical selection rules for two polaron branches (*33, 35*).) Following the optical selection rules discussed above, the spontaneous spin (valley) polarization is aligned (anti-aligned) in two layers (Fig. 1E). The assignment is further supported by the nearly identical magnetic-field dependence in two layers. The attractive polaron responds mostly to the $\sigma^+$ excitation for $B < 0$ T and to the $\sigma^-$ excitation for $B > 0$; except an abrupt change around $B = 0$, the attractive polaron response varies monotonically with field and saturates at high fields. Note that if the spontaneous spin polarization were anti-aligned (Fig. 1D), a switch of the excitation helicity for one of the layers would be expected before saturation because all spins should be aligned after magnetic saturation; this is in contradiction to our experimental observation.

**Signature of a canted spin texture**
We summarize the magnetic-field dependence of the out-of-plane spin polarization by following the maximum MCD of the attractive polaron feature in each layer (Fig. 4A). The low-field behavior including the magnetic hysteresis is included as an inset. The sign of MCD depends on the multilayer interference effect but the attractive polaron oscillator strength for each helicity does not (Fig. 3A). We thus set MCD in both layers to be positive for $B > 0$ since the spin polarization in two layers is aligned. For comparison, we also normalize MCD in each layer to its saturated value at 8 T. Beyond the hysteresis loop, MCD increases monotonically with magnetic field in both layers (Fig. 4A). In $WSe_2$, it saturates near 6 T, and the zero-field MCD is about 60% of the saturation value. In $MoTe_2$, MCD saturates near 8 T, and the zero-field value is about 45% of the saturation value. This behavior is qualitatively different from that of $R_{xy}$, which is quantized at zero field and does not depend on the magnetic field. Our result shows that full spin polarization is not required for quantized Hall transport; this is also consistent with the absence of a MCD plateau even when there is a quantized $R_{xy}$ plateau in Fig. 2C.

We also perform temperature dependence study of MCD in Fig. 4B to elucidate the role of thermal fluctuations. All curves are normalized by the MCD value at 8 T and 1.6 K in each layer (i.e. magnetic saturation). As temperature $T$ increases, the spontaneous MCD weakens, but the slow magnetic saturation beyond the hysteresis loop remains nearly unchanged. Figure 4C (top) displays the temperature dependence of the spontaneous MCD (the spectra are included in Supplementary Fig. 4). It decreases monotonically with increasing temperature; for $T \leq 2K$, the change is small, indicating saturated spontaneous polarization. Above the magnetic ordering temperature (5 - 6 K), the spontaneous MCD decreases to zero. We can also extract the slope of the magnetic-field dependence of MCD at zero field, which is proportional to the spin or magnetic susceptibility, $\chi$ (*36*). The temperature dependence of $\chi$ is well described by the Curie-Weiss law, $\chi^{-1} \propto T - \theta$, with Curie-Weiss temperature $\theta \approx 5$ K (dashed line, Fig. 4C bottom panel). The value reflects the energy scale of the ferromagnetic interaction between the local moments in the QAH insulator. Thermal fluctuations at 1.6 K are therefore unimportant. Our result supports canted spin with nonzero in-plane component in the QAH state.

**Discussions**
The spin-polarized QAH ground state observed in this experiment is unexpected. The current theoretical studies without considering the spin-dependent interactions have shown that the valley-polarized state is the stable ground state (Fig. 1D), i.e. hybridization of states from different layers but from the same valley near the Fermi level (*24-29*). The spin-polarized QAH state requires mixing states from different layers and different valleys; a correlated charge gap is opened at the Fermi level; the state is thus spontaneous layer- and valley-coherent (*30, 31*). Such a state could occur, for instance, via an exciton condensation mechanism near band inversion (*30, 31*). The excitonic mechanism could be important here given the weak interlayer coupling in AB-stacked

TMD bilayers (*29-31*). Future experiments with an ultrathin spacer between the TMD bilayers that quenches the interlayer tunneling while maintaining the strong Coulomb interaction and reasonable moiré potential could test the validity of the mechanism.

The origin of the observed canted spin texture for the QAH state is also not understood. One possible explanation comes from the close proximity of the QAH insulator to a 120°-Néel-ordered Mott insulator (an intralayer- and intervalley-coherent state) and the possibility of inheriting the noncollinear spin configuration (*25*). The weaker spontaneous MCD and higher saturation magnetic field in MoTe$_2$ could be related to the weaker Ising spin-orbit coupling in MoTe$_2$. Another possibility is the Rashba spin-orbit coupling, which becomes important under high out-of-plane electric fields and can cause spin canting. Our results call for a better understanding of the microscopic origin of the QAH state and its magnetic ground state in AB-stacked MoTe$_2$/WSe$_2$ moiré bilayers.

**Acknowledgements**
We thank Yang Zhang, Trithep Devakul, Liang Fu, Yongxin Zeng, Nemin Wei, Allan MacDonald, Ming Xie, K. T. Law, Ya-Hui Zhang and Feng Wang for many insightful discussions.

**Figures**

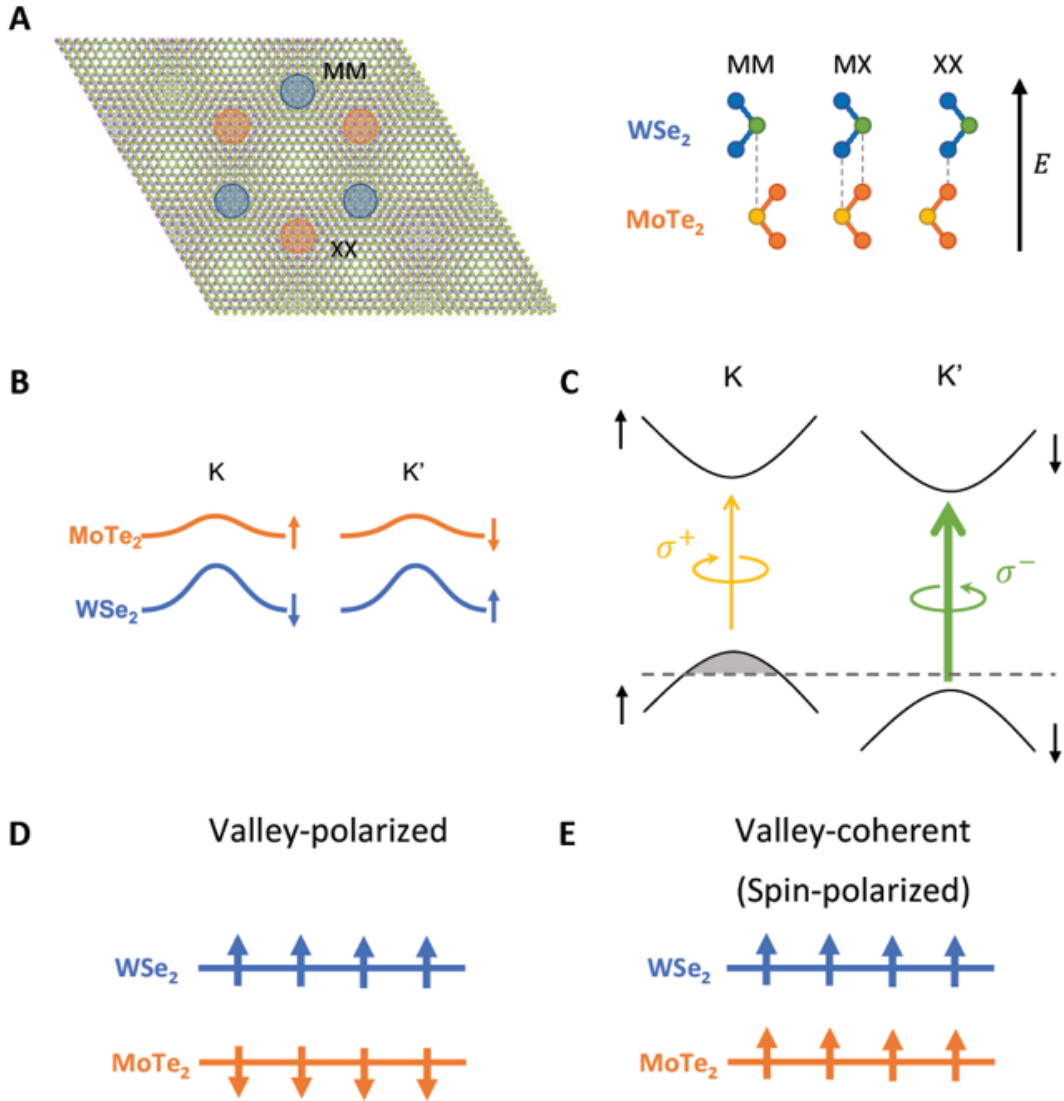

**Figure 1 | AB-stacked MoTe$_2$/WSe$_2$ moiré bilayers. A,** Moiré pattern (left) formed in 60°-aligned (or antiparallel) MoTe$_2$ and WSe$_2$ bilayers (right). MM, MX and XX (M = Mo, W; X = Se, Te) are the high-symmetry stacking sites. The Wannier orbital of the two layers occupies the MM and XX sites, forming a honeycomb lattice. An out-of-plane electric field ($E$) from MoTe$_2$ to WSe$_2$ induces the QAH effect. **B,** Schematics of the topmost valence bands at the K and K' valley from two layers. Arrows denote the out-of-plane spin alignment. In each layer, spin and valley are locked. In each valley, two bands have opposite spins. **C,** Optical selection rules in monolayer TMDs. Attractive polaron, a bound state of the electron-hole excitation in the K (K') valley and hole excitations in the K' (K) valley, couples exclusively to the $\sigma^+$ ($\sigma^-$) excitation. When the holes occupy K valley only, the $\sigma^-$ excitation dominates. The dashed line denotes the Fermi level. **D,E,** Two possible QAH ground states: valley-polarized (**D**) and valley-coherent (or spin-polarized) (**E**), in which the spin polarization is anti-aligned and aligned in two layers, respectively.

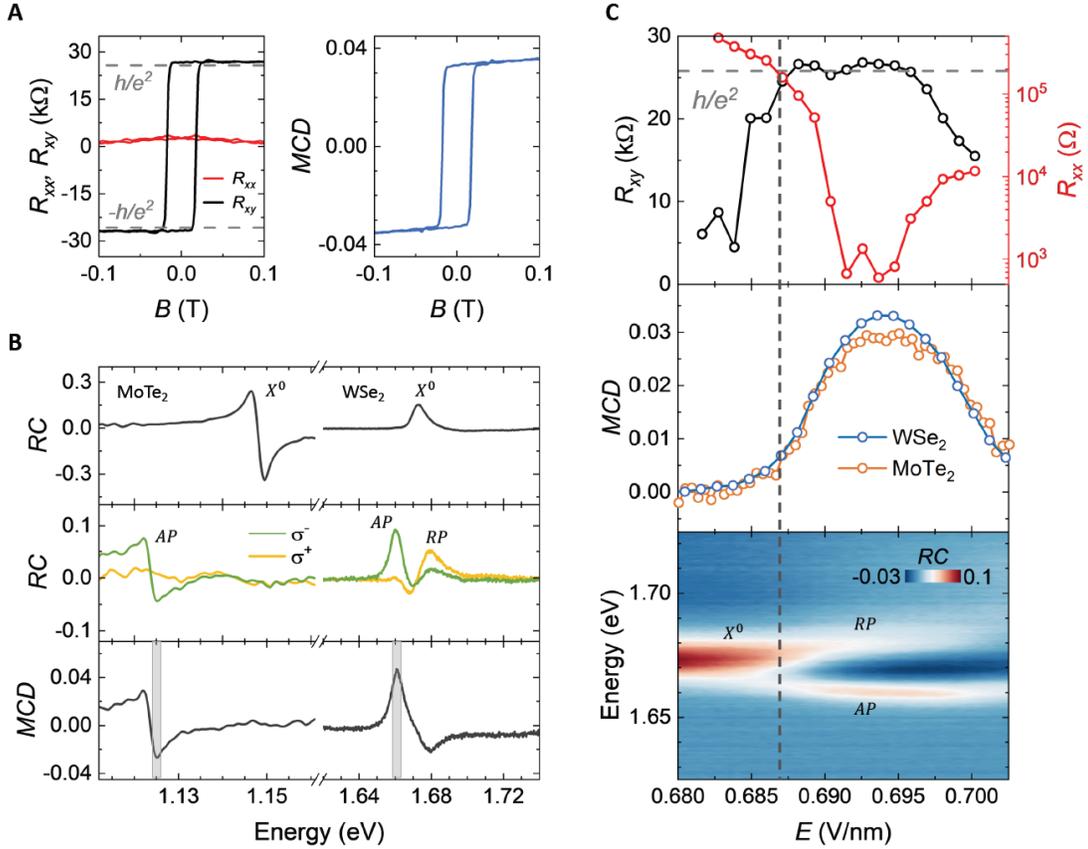

**Figure 2 | QAH effect and topological quantum phase transition at 1.6 K. A**, Hysteretic magnetic field dependence of $R_{xx}$, $R_{xy}$ (left) and MCD in WSe$_2$ (right) with fixed $E \approx 0.693$ V/nm and doping density of one hole per moiré unit cell. Quantized $R_{xy}$ at the value of $h/e^2$ (dashed lines) and vanishing $R_{xx}$ at zero magnetic field support the QAH effect. **B**, Helicity-resolved reflectance contrast (RC) spectrum of the device in the charge neutral state (top) and the QAH state under zero magnetic field (middle). The two helicities are degenerate in the charge-neutral state. The MCD spectrum (bottom) is extracted from the data in the QAH state. The low- and high-energy spectra are dominated by MoTe$_2$ and WSe$_2$, respectively. X$^0$, AP, RP denote the neutral exciton, attractive and repulsive polarons, respectively. **C,** Electric-field dependence of $R_{xy}$, $R_{xx}$ (top), absolute MCD in two layers (middle) and RC spectrum of WSe$_2$ (bottom), all under zero magnetic field. Vertical dashed line denotes the critical field $E_c \approx 0.687$ V/nm for the topological transition from a Mott insulator to a QAH insulator. It also marks the onset of charge transfer from MoTe$_2$ to WSe$_2$.

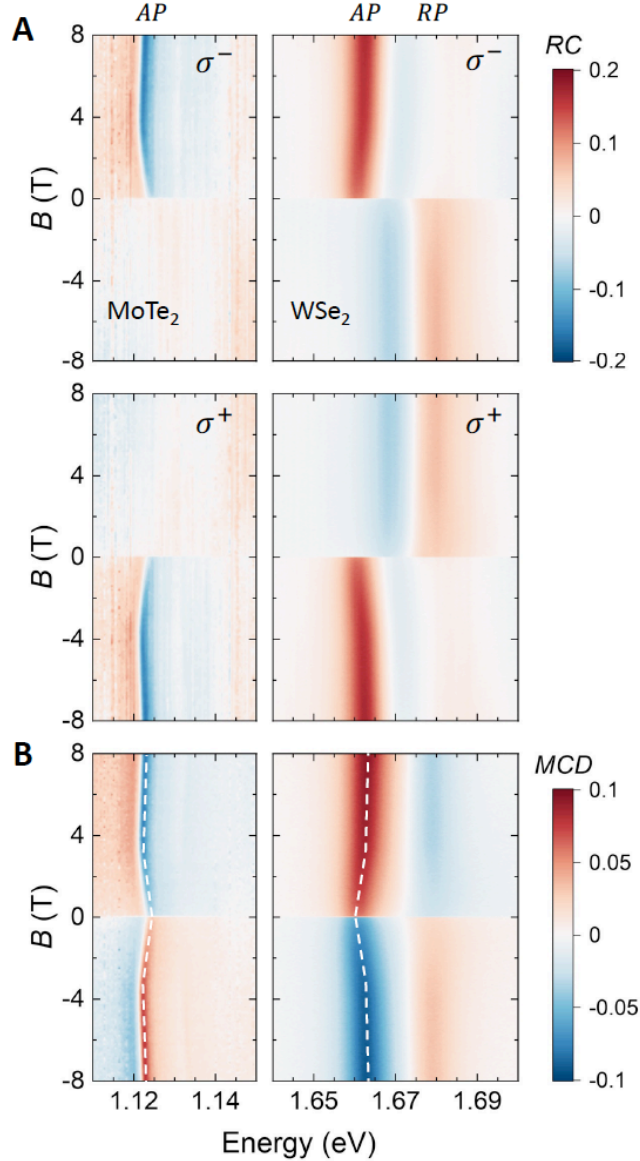

**Figure 3 | Spontaneous spin alignment in the bilayer. A,** Magnetic-field dependence of the right-handed ($\sigma^-$, top) and left-handed ($\sigma^+$, bottom) reflectance contrast spectrum of MoTe$_2$ (left) and WSe$_2$ (right) of the bilayer in the QAH state at 1.6 K. The electric field is fixed at 0.693 V/nm. The two helicity-resolved spectra were measured together at each magnetic field while it was scanned from - 8 T to 8 T. The doping density was adjusted at each field following the Streda relation to stay in the QAH state. The two layers have nearly identical magnetic-field dependence. **B,** Magnetic-field dependent MCD spectra for MoTe$_2$ (left) and WSe$_2$ (right) extracted from the data in **A**. The white dashed lines denote the maximum (absolute) value of the MCD in each layer.

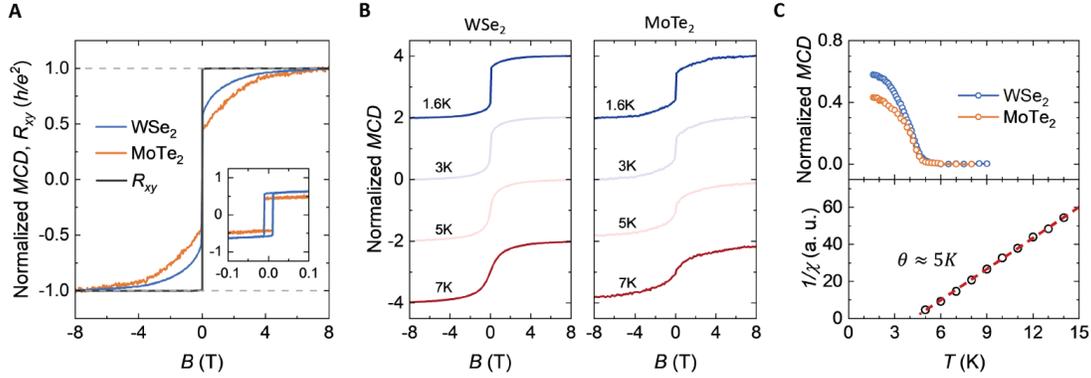

**Figure 4 | Signature of a canted spin texture. A,** Magnetic-field dependence of normalized MCD in each layer in the range of ± 8 T and ± 0.1 T (inset) at 1.6 K. The MCD is averaged over a narrow spectral range of 0.9 meV centered at the maximum of the attractive polaron feature (dashed lines in Fig. 3B). It is normalized by the value at 8 T in each layer. The MCD shows an abrupt change near zero field and saturates slowly with increasing field. In contrast, $R_{xy}$ (right axis) is quantized at the value of $h/e^2$ (dashed lines) near zero magnetic field and does not depend on field. **B,** Same as **A** at varying temperatures. All curves are normalized by MCD at 8 T and 1.6 K in each layer. The curves are vertically displaced for clarity. **C,** Temperature dependence of the spontaneous MCD for MoTe$_2$ and WSe$_2$ (top). Both saturates below 2 K and vanishes above 5 - 6 K. Bottom: Temperature dependence of the spin (or magnetic) susceptibility in WSe$_2$ (symbols) above the magnetic ordering temperature. It is well described by the Curie-Weiss law, $\chi^{-1} \propto T - \theta$, with a Curie-Weiss temperature $\theta \approx 5$ K (red line).

# Supplementary Materials for
## "Valley-coherent quantum anomalous Hall state in AB-stacked MoTe$_2$/WSe$_2$ bilayers"

Zui Tao[*], Bowen Shen[*], Shengwei Jiang[*], Tingxin Li[*], Lizhong Li, Liguo Ma, Jenny Hu, Kateryna Pistunova, Kenji Watanabe, Takashi Taniguchi, Tony F. Heinz, Kin Fai Mak[**], and Jie Shan[**]

[*]These authors contributed equally.
[**]Email: jie.shan@cornell.edu; kinfai.mak@cornell.edu

## Methods
### Device fabrication and electrical characterization

The lattice mismatch between MoTe$_2$ and WSe$_2$ is about 7%. It gives rise to a moiré density of about $5 \times 10^{12}$ cm$^{-2}$ in parallel (AA-stacked) or anti-parallel (AB-stacked) heterobilayers (*1, 2*). Since the density weakly depends on the twist angle for small angles, these moiré structures are generally more homogeneous than twisted homobilayers (*3*). The high moiré density also favors the formation of good electrical contacts for transport studies. We investigate the QAH effect in AB-stacked MoTe$_2$/WSe$_2$ bilayers. They have a type-I band alignment (*2, 4*). The topmost moiré valence band is from the MoTe$_2$ K/K' states (Fig. 1B). Under a large electric field, the topmost moiré valence band from the WSe$_2$ K/K' states can be tuned to overlap with the MoTe$_2$ band (*4, 5*). The QAH effect is the result of electric-field-tuned band inversion and mixing.

We fabricated dual-gated devices of AB-stacked MoTe$_2$/WSe$_2$ moiré bilayers using the layer-by-layer dry transfer method as detailed in Ref. (*2, 6*). All bulk TMD crystals were acquired from HQ Graphene. The crystal orientations of MoTe$_2$ and WSe$_2$ monolayers were determined by the angle-resolved optical second-harmonic generation (SHG) before stacking (*3*). The twist angle in the stacked bilayers was verified also by SHG. The angle alignment is typically within $\pm\,0.5°$ of the target angle. The moiré bilayers are imbedded in two gates; both are made of hexagonal boron nitride (hBN) gate dielectrics and few-layer graphite gate electrodes. The thickness of hBN in the top and bottom gates is about 5 and 20 nm, respectively. The two gate voltages combined tune the out-of-plane electric field and the doping density in the moiré bilayer.

We shaped the devices into Hall bar geometry. The contact design and the transport measurement method are detailed in Ref. (*1, 2*). In short, to achieve good electrical contacts to the TMD bilayers, 5-nm-thick Pt is used as metal electrodes, and the contact region of the TMD bilayers is heavily hole-doped. Both the longitudinal and transverse resistance were measured using the low-frequency lock-in technique with a small bias voltage (1-2 mV). The raw $R_{xx}$ and $R_{xy}$ data under magnetic field $B$ and $-B$ were symmetrized and anti-symmetrized, respectively, to remove the small longitudinal-transverse coupling in our devices. Electrical transport and layer-resolved optical spectroscopy were measured and correlated on one device in several thermal cycles. All results are reproducible. The results were further verified on another device by performing optical measurements on the WSe$_2$ layer.

**Helicity-resolved reflectance contrast spectroscopy**
The optical measurements were performed in a closed-cycle helium cryostat equipped with a superconducting magnet (Attocube, Attodry 2100, base temperature 1.6 K). A broadband tungsten halogen lamp was used as the light source. The output of the lamp was collected by a multimode fiber and focused onto the device under normal incidence using a microscope objective (0.8 numerical aperture). A combination of a linear polarizer and an achromatic quarter-wave plate was used to generate circularly polarized light. The reflected light of a given helicity was spectrally resolved by a spectrometer coupled to a thermoelectric-cooled CCD camera and a thermoelectric-cooled InGaAs one-dimensional array sensor (which capture the intralayer excitations in WSe$_2$ and MoTe$_2$, respectively). The reflectance contrast spectrum was obtained by comparing the reflected light spectrum from the sample to the reference spectrum measured on heavily doped devices (which is featureless in the interested spectral regions). In these measurements, the beam size on the device was about 3.5 $\mu$m in diameter. The incident light intensity was kept below 5 nW/$\mu$m$^2$ to avoid heating or photo-doping by filtering out the spectrum outside the regions of interest.

The MCD spectrum is defined as $\frac{I^- - I^+}{I^- + I^+}$, where $I^-$ and $I^+$ denote the reflection intensity of the $\sigma^-$ and $\sigma^+$ light, respectively. The MCD is strongly enhanced near the attractive polaron resonance in each layer, and is used to characterize the spin polarization in the corresponding layer. Because the attractive polaron resonance shifts in energy with magnetic field (Fig. 3), we follow the MCD maximum and average the value in a narrow spectral window of 0.9 meV around the maximum to obtain the result in Fig. 4. We set the MCD in both layers to be positive for $B > 0$ and negative for $B < 0$ since the spin polarization is aligned in two layers.

We employed a laser-based lock-in measurement technique to improve the signal-to-noise ratio of the magnetic susceptibility in WSe$_2$ (Fig. 4C bottom panel). A continuous-wave Ti-sapphire laser (M Squared SOLSTIS system) was used as the light source. We chose a fixed wavelength (747.4 nm) near the attractive polaron resonance, which has a negligible temperature dependence below 15 K. The incident light power was limited to about 100 nW on the device to minimize heating. The light helicity was modulated by a photo-elastic modulator at 50.1 kHz and the reflected light was detected by a photodiode. The MCD signal is defined as the ratio of the modulated signal (measured by a lock-in amplifier) to the total signal (measured by a DC voltmeter). The magnetic susceptibility was extracted from the magnetic-field dependence of MCD in a small field range up to 0.1 T.

**Supplementary References**

**Supplementary Figures**

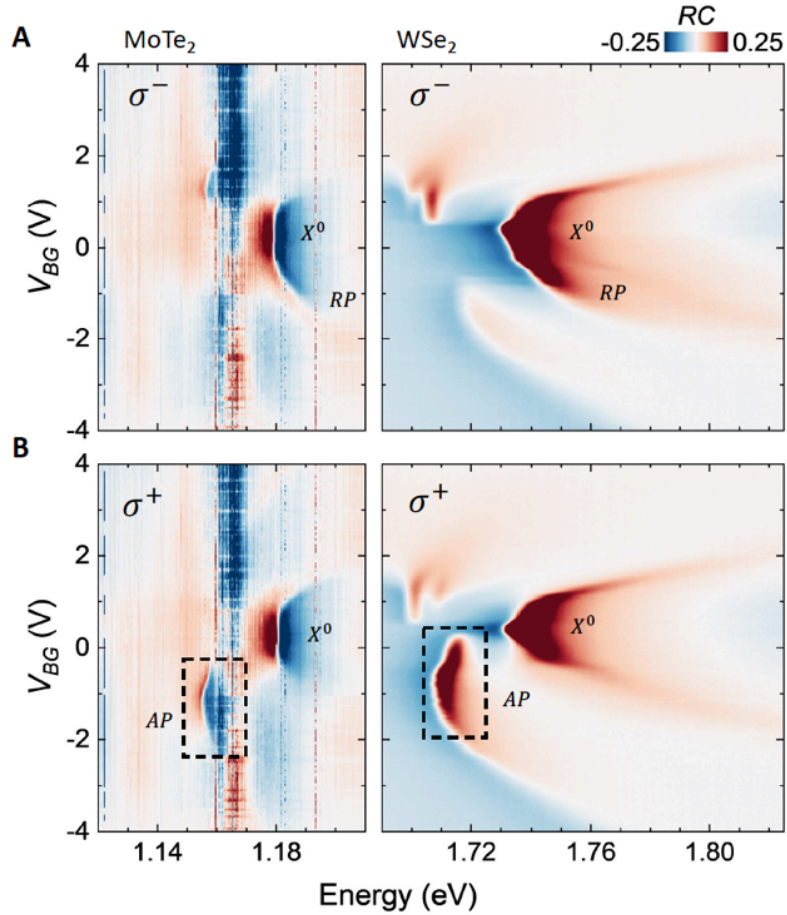

**Supplementary Figure 1 | Optical selection rules in TMD monolayers. A,B,** Helicity-resolved reflectance contrast spectrum of MoTe$_2$ and WSe$_2$ monolayers as a function of back gate voltage, $V_{BG}$, under a constant out-of-plane magnetic field of - 5 T. The optical excitation is right ($\sigma^-$) (**A**) and left ($\sigma^+$) (**B**) circularly polarized. The two monolayers are on the same substrate next to each other. The gate voltage induces electron doping ($V_{BG} >$ 0) or hole doping ($V_{BG} <$ 0) in the TMDs. The spins are aligned by the applied magnetic field in both layers. The two layers follow the same optical selection rules for hole doping: the attractive polaron resonance in both layers is strong in the $\sigma^+$ channel and weak in the $\sigma^-$ channel.

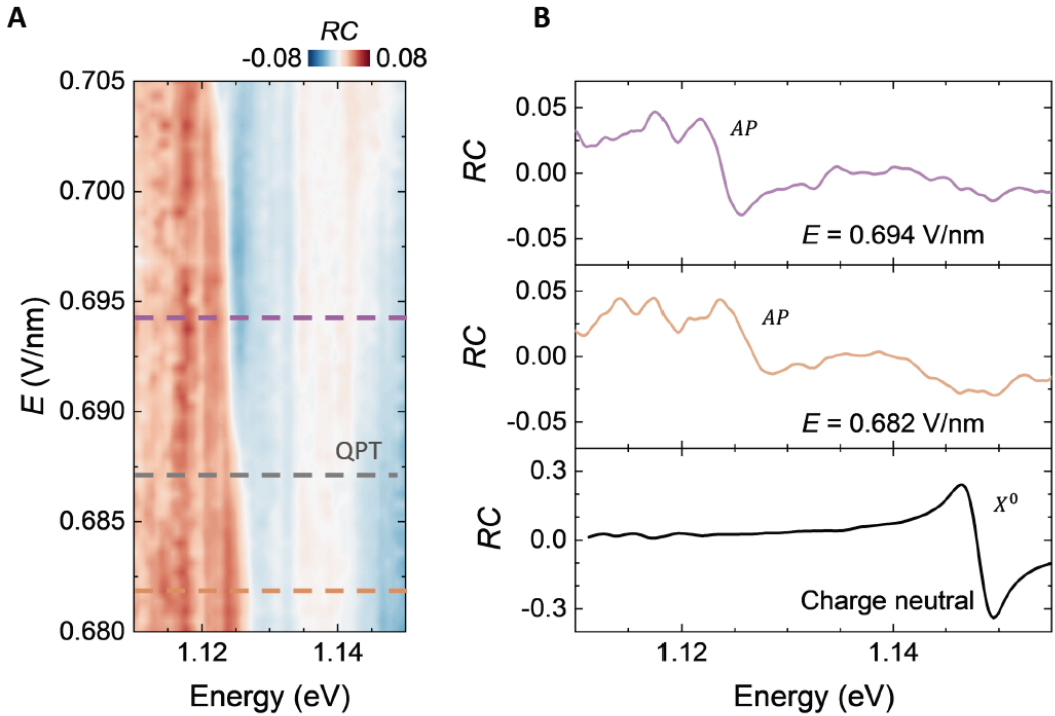

**Supplementary Figure 2 | Reflectance contrast of MoTe$_2$ across the topological quantum phase transition. A,** Un-polarized reflectance contrast spectrum of MoTe$_2$ at 1.6 K as a function of out-of-plane electric field. The black dashed line denotes the critical field for the topological quantum phase transition as determined in Fig. 2. **B,** Linecuts at two fields (top and middle), one from each phase, show that the spectral response is dominated by the attractive polaron, and MoTe$_2$ is doped for the entire electric-field range in **A**. The response of charge neutral MoTe$_2$ is included as a reference (bottom).

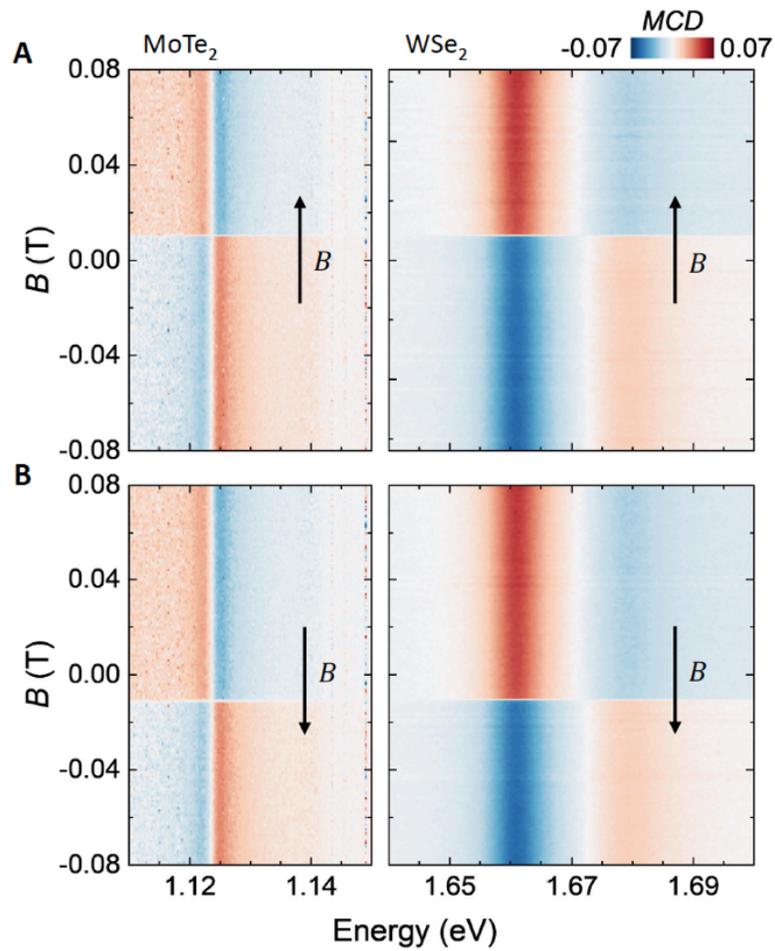

**Supplementary Figure 3 | MCD spectra for magnetic-field range of ± 0.08 T. A,B,** MCD spectrum as a function of magnetic field between - 0.08 T to 0.08 T in the forward (**A**) and backward (**B**) scan directions. The low- and high-energy spectra are dominated by the response of MoTe$_2$ and WSe$_2$, respectively. Hysteric behavior is observed in both layers.

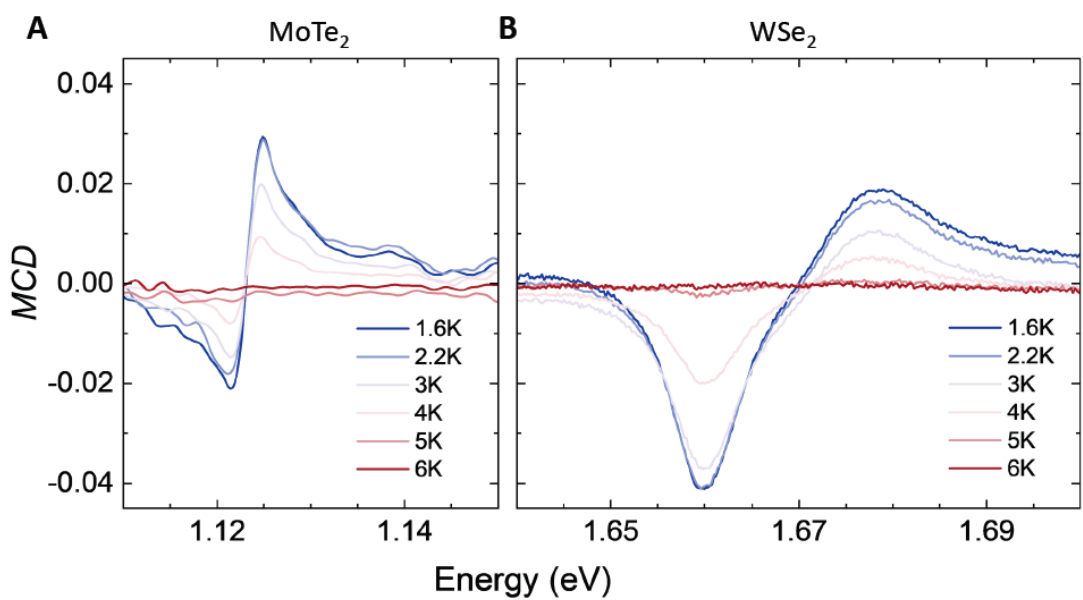

**Supplementary Figure 4 | Temperature dependence of spontaneous MCD spectrum.**
**A,B,** MCD spectrum under zero magnetic field for MoTe$_2$ (**A**) and WSe$_2$ (**B**) at varying temperatures. The magnitude decreases monotonically with increasing temperature and vanishes between 5 - 6 K.